\documentclass[11pt]{article}
\usepackage{Blois,epsfig}

\usepackage{amssymb}

\usepackage[super,compress]{cite}
\usepackage{graphicx}
\def\pom{$I\!\!P \:$}
\def\dpe{$\mathrm{DI}\!\!\mathrm{PE} \:
$}
\begin{document}
\begin{flushright}
FERMILAB-PUB--14-270E
\end{flushright}
\vspace*{1cm}

\vspace*{2cm}
\title{Central Exclusive Production at the Tevatron}



\author{Michael Albrow \\
On behalf of the CDF Collaboration}

\address{Fermi National Accelerator Laboratory\\
P.O.Box 500, Wilson Road, Batavia, IL 60510, USA\\
albrow@fnal.gov}

\maketitle

\begin{abstract}
The Collider Detector at Fermilab, CDF, at the Fermilab Tevatron, observed for the first time in hadron-hadron ($p\bar{p}$) collisions
exclusive two-photon production, as well as
photon-photon ($\gamma + \gamma \rightarrow e^+e^-, \mu^+\mu^-$) and photon-pomeron ($\gamma + $\pom$ \rightarrow J/\psi, \psi(2S)$)
interactions, and $p+\bar{p} \rightarrow p + \chi_c + \bar{p}$ by double pomeron exchange, \pom + \pom $\;$ or \dpe.
Exclusive $\pi^+\pi^-$ production was also measured at $\sqrt{s}$ = 900 GeV and 1960 GeV; $f_0$ and $f_2$ resonance structures are discussed.

\end{abstract}

\section{Introduction}	

In high energy hadron-hadron collisions there are interactions that are inelastic, but have two leading protons with some features of elastic scattering. 
We may write $p + p \rightarrow p + X + p$, where the $p$ stands for proton or antiproton, and the
produced particles (generically termed $X$) are in the ``central region" in rapidity $y$\cite{kine}, separated by large rapidity gaps $\Delta y$ from the
quasi-elastically scattered (anti-)protons, which may or may not diffractively dissociate into low-mass clusters.
These reactions are called ``Central Exclusive Production" (CEP). At high energy the types of colliding hadrons should be irrelevant; $X$ is formed from exchanged
photons or
pomerons \pom, so there are three classes: $\gamma\gamma \rightarrow X, \gamma$ \pom $\rightarrow X$, and \pom\pom $\rightarrow X$.
The pomeron is a strongly interacting color-singlet with vacuum quantum numbers, at leading order a pair of gluons, but as a $t$-channel exchange it
is not a real particle.
The two-photon reactions were studied in great detail at $e^+e^-$ colliders. At the high
energies of an ILC or CLIC nearly all of the total inelastic cross section is not annihilation but $\gamma\gamma \rightarrow X$. 
With $e+p$ collisions at HERA measurements were made of $\gamma$ \pom
reactions (photoproduction). The photon fluctuates to a vector meson, or a quark loop, which scatters by \pom -exchange on the proton. 
Photoproduction of $J/\psi (c\bar{c})$ or $\Upsilon (b\bar{b})$ can be calculated semi-perturbatively by two-gluon exchange between the quark loop and the proton. 
For light vector mesons such as $\rho$ and $\phi$ perturbative QCD cannot be applied, and the treatment is phenomenological. 
 The third class, \pom \pom, is unique to hadron-hadron collisions and dominates over the other two classes. Thus CEP probes
an aspect of QCD that is not fully understood (i.e. ``calculable").  CDF observed $\gamma\gamma$ and $\gamma$\pom interactions 
in hadron-hadron collisions for the first time, as well as \pom \pom $\rightarrow \gamma\gamma, /; \chi_c$.

\begin{figure}[!ht]
  \begin{center}
     \includegraphics[width=0.95\textwidth]{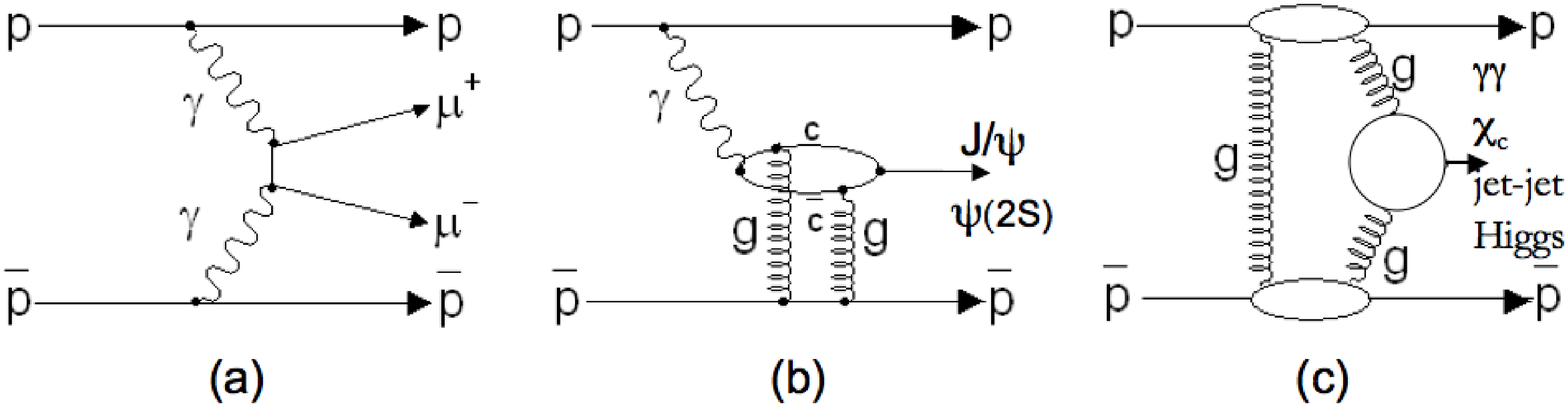}
    \caption{(a) Two-photon production of lepton pairs (b) Photoproduction of vector charmonium (c) Hard double pomeron processes.
    $\gamma\gamma$ is mainly through u- and c-quark loops, $\chi_c$ though a $c$-loop and $H$ through a top-quark loop. Exclusive
    di-jets are produced without quark loops.}
  \end{center}
  \label{threediag}
\end{figure}

Figure \ref{threediag} shows diagrams for the three classes.

The Tevatron, with $\sqrt{s}$ up to 1960 GeV, had a much higher c.m. energy than earlier
fixed target experiments\cite{kirk,gutierrez} ($\sqrt{s} < 40$ GeV) and the Intersecting Storage Rings (ISR) 
experiments\cite{albrowisr,makariev} ($\sqrt{s} <$ 63 GeV). This allowed much higher central masses $M(X)$ in $p + X + p(\bar{p})$
reactions with two rapidity gaps\cite{kine} $\Delta y >$ 3. In addition, for small $M(X) \lesssim$ 4 GeV (in the resonance region) one can have much larger
rapidity gaps $\Delta y$, with correspondingly greater purity for reactions with $t$-channel exchanged photons, $\gamma$, or pomerons, \pom.
 Other exchanged reggeons, such as the $\rho$-trajectory, are exponentially suppressed in $\Delta y$. The Tevatron energy has now been exceeded by a factor $\sim$4 by the Large Hadron Collider, LHC, allowing correspondingly higher $M(X)$
up to hundreds of GeV, with higher $E_T$ jets and exclusive $W^+W^-$ production possible \cite{cmsww}. Exclusive Higgs boson H(125) production
with no other particles produced is also expected\cite{durhamh} with $\sim$fb cross sections. The observation of exclusive
di-jets by the Tevatron experiments CDF and DO$\!\!\!\!$/ is covered elsewhere in this issue\cite{goulianos}. I shall present results
from CDF on $\gamma + \gamma \rightarrow e^+e^-$ and  $\mu^+\mu^-$, $\gamma + $\pom$ \rightarrow J/\psi, \psi(2S)$, \pom + \pom $\rightarrow \chi_c$,
and \pom + \pom $\rightarrow \pi^+\pi^-$ with mass up to $M(\pi\pi$) = 5 GeV/c$^2$. The latter measurement used data at both $\sqrt{s}$ = 900 GeV and 1960 GeV.  
CDF also made a search for exclusive $Z$ photoproduction. I first describe some generic issues common to the different analyses, and then consider
the specifics of each measurement. For more details see the original papers.

\begin{figure}[!ht]
  \begin{center}
     \includegraphics[width=0.95\textwidth]{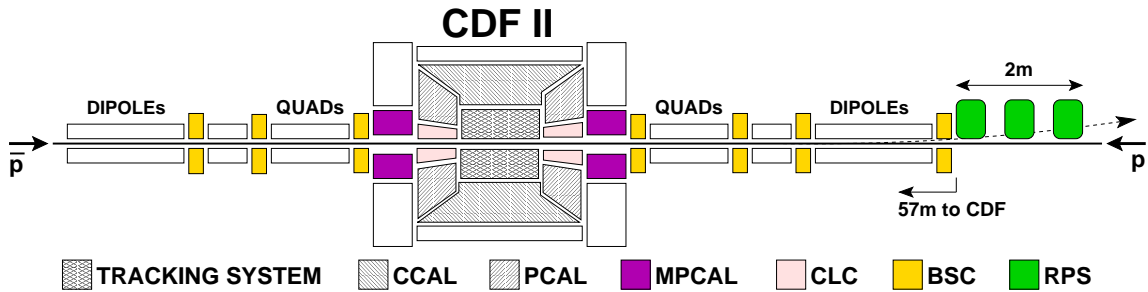}
    \caption{Schematic diagram of CDF II including the forward detectors (not to scale!). CCAL = Central calorimeter, PCAL = Plug calorimeter, 
    MPCAL = Miniplug calorimeter, CLC = Cherenkov luminosity counters, BSC = Beam shower counters, and RPS = Roman pot spectrometer.}
  \end{center}
  \label{cdffwdtoo}
\end{figure}

The data presented here did not have detection of the scattered quasi-elastic (anti-)protons.
Fig. \ref{cdffwdtoo} shows schematically (not to scale) a plan view of the CDF II detector. The proton beam points to the $+z$ (+ve $\eta$)
direction. (This figure is from Ref. \citen{diffwz} on diffractive $W$ and $Z$ production.)
Earlier Roman pots in Run I (at $\sqrt{s}$ = 1800 GeV), with silicon and drift-chamber tracking on each downstream arm and used to measure elastic scattering\cite{cdfelastic},
 were removed before Run II. Therefore the data reported here include events where one or both protons (meaning $p$ or $\bar{p}$) dissociated into a cluster of particles, all with
$|\eta| > |\eta_{max}|$, where $|\eta_{max}|$ was the limit of the CDF detector coverage. The CDF detector is described in Ref.\citen{cdf}. It had
cylindrical silicon strips (not used in these studies) and drift chamber tracking in the central region, $|\eta| < 1.3$, in a 1.4 Tesla solenoidal field. 
The drift chamber had 96 layers of wires, axial and $\pm 3^\circ$ stereo, with 180 $\mu$m hit measurement precision from the drift time, and pulse-height measurement
providing some particle identification by dE/dx ionization.
A barrel of
time-of-flight counters (ToF) was outside the tracking, and had  $279 \times 4 \times 4$ cm$^3$   scintillator bars with a Hamamatsu R7761 photomultiplier (PMT) on each end, 
giving a time resolution of $\sim$120 ps. This was surrounded by 
electromagnetic (EM) lead-scintillator and hadronic (HAD) iron-scintillator calorimeters covering $|\eta| < 3.64$, 
a ``Miniplug" lead-liquid scintillator calorimeter ($3.6 \lesssim |\eta| \lesssim 5.2$)
and a set of very forward ``beam shower counters" (BSC), which were scintillator plates with PMTs. The latter were used as ``rapidity gap detectors", detecting showers produced
by large $|\eta|$ particles hitting the beam pipes or surrounding material. For the earlier results\cite{cdfee,cdfmumu,cdfz,cdfgg1,cdfgg} very forward 
coverage was obtained
using three BSC stations, at $|z|$ = 6.6 m, 23.2 m, and 31.6 m. These covered $|\eta| \lesssim 7.4$. Since there were quadrupole
magnets in front of BSC-2 the $|\eta|$-coverage depended on the charge and momenta of the forward particles. BSC-2 and BSC-3 were rectangular
and so the polar angle coverage was $\phi$-dependent. The beam (true) rapidity was
$y_{BEAM} = \mathrm{ln} \frac{\sqrt{s}}{m_p} = 7.64 (6.87)$ at $\sqrt{s}$ = 1960 (900) GeV, which allowed very little rapidity for undetected dissociation products.
CDF estimated the fraction of final events with undetected diffraction dissociation to be $\lesssim$8\% in the two-photon mediated lepton pair events, and $<$ 1\% in
exclusive di-photon production by \dpe.

For the most recent results only the closest station (BSC-1) could be used.
BSC-1, covering $5.4 \lesssim |\eta| \lesssim 5.9$, had no transverse 
magnetic field upstream and it was preceded by two radiation lengths of lead to convert photons. There were four PMTs covering four azimuthal quadrants. 
The ``OR" of these counters was put in VETO in the level-one trigger for these studies, reducing pile-up from more than one
interaction in the same bunch crossing, as well as excluding a large fraction of the inelastic collisions, while retaining the CEP reactions.
The mean number of interactions per bunch crossing, $\bar{\mu}$, was as high as eight at the beginning of the later stores,
so the CEP trigger was only active when and if $\bar{\mu}$ had decreased to $\lesssim$ 3. All cross-section measurements were made with no additional
inelastic collisions, and CDF defined the \emph{effective} luminosity $L_{eff}$ as the integrated luminosity taking that into account. Since the
(36) colliding bunches did not all have the same $\bar{\mu}$ the calculation was made bunch-by-bunch. As the maximum rapidity coverage of CDF with BSC-1 was
$|\eta|_{max}$ = 5.9, higher mass dissociations were included, and more so at $\sqrt{s}$ = 1960 GeV than at 900 GeV. All data reported here are at $\sqrt{s}$ = 1960 GeV 
except for some $X = \pi\pi$ data\cite{cdfpipi}.

Both forward regions were also equipped with Cherenkov Luminosity Counters (CLC), consisting of 48 gas (isobutane) tubes pointing to the interaction region, each
viewed by a PMT. These arrays covered $3.7 < |\eta| < 4.7$, with 16 units in each of three $|\eta|$ layers. In addition to monitoring the
bunch-by-bunch luminosity, $L_{bunch}$, they were included in the rapidity-gap selection.

The use of zero-bias (bunch crossing) events to define an ``empty detector" when there was no inelastic collision
with any particles in the acceptance ($|\eta| < 7.5$ or 5.9) was very important for these analyses. Zero-bias events were recorded at a rate of about 1 s$^{-1}$
throughout the data taking. The zero-bias data was divided into two classes of events: (1) No tracks, no CLC hits and no muon stubs, classed
as ``No interaction", and (2) All other events, classed as (one or more) ``Interaction". CDF then plotted the summed pulse height distributions for
each detector, namely the central EM and
HAD calorimeters ($|\eta| <$ 1.3), the Plug calorimeter ($1.32 < |\eta| < 3.64$), the Miniplug calorimeter and the CLC and BSC counters.
The ``No interaction" distributions show the summed noise, and a cut was applied that includes $\sim$99\% of empty events in each subdetector. The ``Interaction" events
are mostly well separated in pulse height, with much larger signals, apart from a small noise peak corresponding to single interactions with 
no particles in a particular subdetector.
One analysis used the pulse height distributions of the PMT in each subdetector with the highest signal (the ``\emph{hottest}" PMT)
instead of, or in addition to, the sum. Thus while the cut on the total energy in the
central calorimeters (EM + HAD) was 2.8 GeV, the requirement on the hottest PMT was 80 (200) MeV for the EM (HAD) calorimeters. The values quoted are examples;
they were optimized for each analysis (the data were taken over eight years). 

With the chosen cuts defining ``No interaction" in the full detector, CDF then plotted the probability P(0) that the full detector was empty vs. $L_{bunch}$, which should be an
exponential since $\mu$ is a Poisson distribution, with mean $\bar{\mu} = \sigma_{vis} \times L_{bunch}$. The exponential fits were good, 
with intercepts P(0)$ \gtrsim 0.99$
at $L_{bunch} = 0$ , confirming that the full detector is classed as empty when there are no collisions. The slope is $(-) \sigma_{vis}. L_{bunch}$,
allowing a measurement of the inelastic cross section by taking the luminosity from the calibrated (to $\pm$6\%) CLC counters and correcting for the ratio
$f_{vis} = \sigma_{vis}/\sigma_{inel}$. The $X = \pi\pi$ data at 1960 GeV is in agreement with $\sigma_{inel} = (61.0\pm1.8)$ mb 
from a global fit including \textsc{totem} values \cite{totem},
and using $f_{vis} = 0.85\pm0.05$ as expected from the Minimum-Bias Rockefeller (MBR) event generator\cite{mbr}. For the corresponding 900 GeV data the CLC counters were not calibrated
and CDF used the predicted $\sigma_{vis}$ = (47.4$\pm$3.0) mb with $f_{vis} = 0.90\pm0.05$ to normalize the cross sections.

For the $p + \bar{p} \rightarrow p(*) + X + \bar{p}(*)$ physics events CDF used the same ``empty detector" criteria, just excluding the $X$ decay products and the surrounding calorimeter clusters. In some analyses
the noise cuts were made on the pulse-height sums, in some on the hottest PMT, and in some cases on both; see the original papers for details.

   Table I gives a summary of the results.
 
\section{$X = e^+e^-$}

Motivated by a search for exclusive two-photon production (see Section 4) CDF took data\cite{cdfee} with a trigger on two EM clusters
in $|\eta| <$ 2.0 with $E_T >$ 5 GeV, with BSC-1 in veto. After all the exclusivity cuts there were 16 events with
exactly two oppositely-charged tracks pointing to clusters with HAD:EM ratios and shower shapes consistent with those of electrons.
The azimuthal angle difference of all pairs was $(\Delta \phi - \pi) <$ 0.1, consistent with the \textsc{lpair} generator\cite{lpair} prediction together with the CDF detector simulation,
as were the other measured distributions. CDF found 
$\sigma^{E_T > 5 GeV, |\eta| < 2} = 1.6^{+0.5}_{-0.3}$(stat) $\pm$0.3 (syst) pb at $\sqrt{s} = 1960$ GeV, with a $M(e^+e^-$) spectrum spanning from 10 - 40 GeV/c$^2$.
This was the first observation, with 5.4$\sigma$, of two-photon processes in hadron-hadron collisions.

The theoretical uncertainty in the cross section for $p + \bar{p} \rightarrow p + e^+e^- + \bar{p}$, the ``elastic" case, is only $\sim$ 1\%. 
It is a QED process with only small corrections from the electromagnetic
form factor of the proton and from shadowing corrections (i.e. a simultaneous pomeron exchange, small because the impact parameter for photon exchange events is
usually $\gtrsim 1$ fm). In the CDF case, with diffraction dissociation of the protons allowed, the uncertainty is larger ($\sim$ 15\%). 
Agreement of the above measurements with theory gave confidence in the method of selecting these rare exclusive events ($3\times 10^{-7}$ of
$\sigma_{inel}$).

Two-photon collisions have become an important field
at the LHC, where the (quasi-real) photon fluxes extend to $\sqrt{s}_{\gamma\gamma} \sim$ 1 TeV. Exclusive di-lepton production: 
$\gamma\gamma \rightarrow e^+e^-, \mu^+\mu^-$ and $\tau^+\tau^-$, with equal cross sections, can be measured with $M(X) \gtrsim$ 
400 GeV/c$^2$. The exclusive dimuon events can be used to check the momentum scale of forward proton spectrometers, since both proton momenta are  
known from the dimuon kinematics. Even if one proton dissociates, or is not in the acceptance of the forward spectrometer, the momentum of the
other proton is well known. Exclusive $W^+W^-$ production is measurable\cite{cmsww}, certainly in the $\ell^+\ell^- \nu \bar{\nu}$ channels 
(with no other tracks
on the dilepton vertex) and is sensitive to quartic gauge couplings. At $M(X)$ = 500 GeV/c$^2$ the $ \gamma + \gamma \rightarrow W^+W^-$ cross section 
is about two orders-of-magnitude greater
than the $\gamma + \gamma \rightarrow \mu^+\mu^-$ cross section because it has spin $J = 1$ in the $t$-channel; while $J$ = 0 sleptons or squarks have much smaller 
cross sections (and also are now excluded by other searches).

Another CDF result was a search\cite{cdfz} for exclusive $Z$ production, using the $e^+e^-$ and $\mu^+\mu^-$ channels. This is not a $\gamma\gamma$ process,
which is forbidden by the Landau-Yang theorem, but occurs by photoproduction, $\gamma$ + \pom $\rightarrow Z$. This is an interesting ``vertex", 
actually a loop diagram, with all-neutral external electromagnetic, strong and weak lines. The standard model prediction is much too small for the Tevatron,
but is in reach at the LHC. Starting with 10$^5$ inclusive $Z(e^+e^-,\mu^+\mu^-)$ events, and requiring an otherwise empty detector, one  intriguing $X = e^+e^-$ candidate 
had $M(e^+e^-)$ = 92.1 GeV/c$^2$, but there were hits in the BSC counters and the event was rejected for that reason. 
Fewer than 0.01 exclusive $Z$ events were expected in the Standard Model. Eight other events
survived all the exclusivity cuts (including the empty-BSC requirement), with $M(\ell^+\ell^-)$ from 40 - 76 GeV (similar to the highest mass 
event from $e+p$ collisions in HERA).
Their kinematics agreed with the \textsc{lpair} predictions\cite{lpair} after folding in the CDF detector simulation. All events had ($\Delta\phi - 180^\circ) < 0.75^\circ$
and $p_T(\ell^+\ell^-) < $ 1 GeV/c. These conditions (together with no associated tracks) should allow such events to be measured 
 with some pile-up, but that was not done. The same analysis was carried out with 2.2$\times 10^6 \; W \rightarrow \ell \nu$ events, 
 of which three survived all exclusivity
cuts. Since exclusive $W^\pm$ production is forbidden, this is a control on the background from undetected particles.

\section{$X = \mu^+\mu-$}

With the aim of observing exclusive $\chi_{c0} \rightarrow J/\psi + \gamma$ CDF installed a trigger requiring two muons with $p_T >$ 1.4 GeV/c and $|\eta| < 0.6$, with a
veto on BSC-1. The spectrum\cite{cdfmumu}  of the exclusive 402 $\mu^+\mu^-$ candidates with $M(\mu^+\mu^-) \in [3.0,4.0]$ GeV/c$^2$ 
is shown in Fig.2. Cosmic ray background
was negligible after a three-dimensional opening angle cut, and using the ToF barrel to check consistency with outgoing muons. There are three
components: $J/\psi, \psi(2S)$, and a continuum which agrees with the \textsc{lpair} expectation, so any background is very small. This was the first observation
of photoproduction in hadron-hadron collisions. The cross sections $\frac{d\sigma}{dy}|_{y=0}$ are consistent with the predictions,
and the ratio $R = \frac{\psi(2S)}{J/\psi} = 0.14 \pm 0.05$ agrees with the HERA measurement (in $e+p$) of $R = 0.166\pm 0.012$.

\begin{figure}[!ht]
  \begin{center}
 \includegraphics[width=0.90\textwidth]{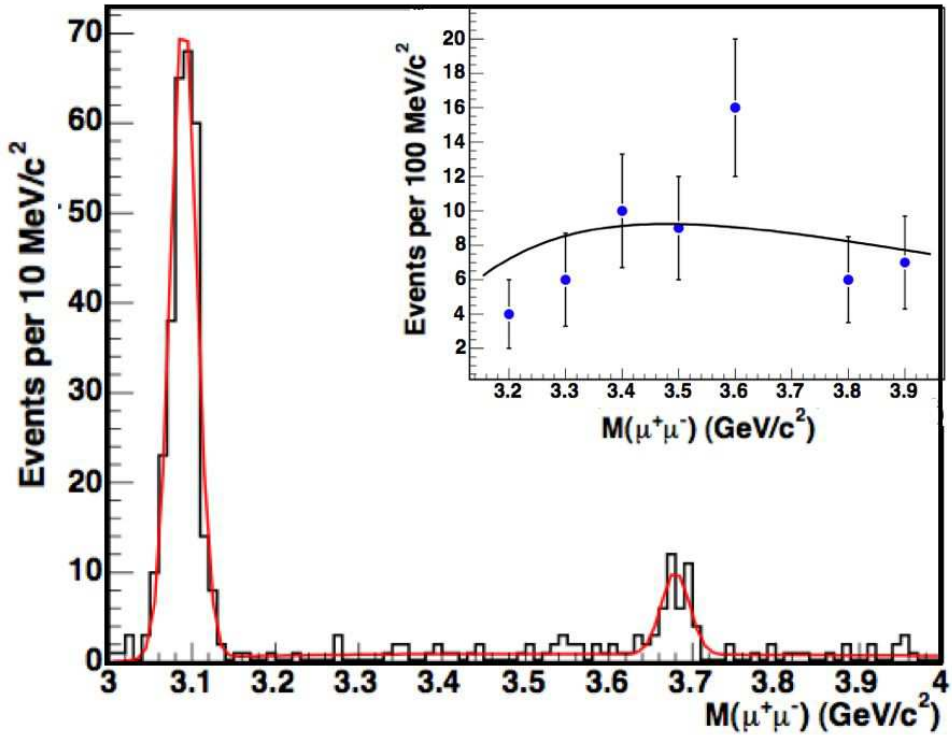}
    \caption{Mass $M_{\mu\mu}$ distribution \cite{cdfmumu} of 402 exclusive events, with no EM shower, (histogram) 
   together with a fit to two Gaussians for $J/\psi$
    and $\psi(2S)$, and a QED continuum. All three shapes are predetermined, with only the normalizations floating. Inset: Data above the 
    $J/\psi$ and excluding $3.65 < M_{\mu\mu} < 3.75$ GeV/c$^2$ ($\psi(2S)$) with the fit to the QED spectrum times
    acceptance (absolute normalization, statistical uncertainties only)(Color online).}
  \end{center}
  \label{mumu34}
\end{figure}

CDF then allowed an additional EM shower with $E_T > 80$ MeV, otherwise keeping the same exclusivity cuts, finding 65 events in the $J/\psi$ region, interpreted as
$\chi_{c,i}^0 \rightarrow J/\psi + \gamma$. As the photons are soft their energy resolution is poor, and CDF could not resolve the three possible $\chi_{c,i}$ states.
The $\chi_{c0}$ is expected to dominate in production, but the $\chi_{c1}$ and $\chi_{c2}$ have higher branching fractions to $J/\psi+\gamma$. 
Assuming that all the events were from $\chi_{c0}$, $\frac{d\sigma}{dy}_{y=0} (\chi_{c0})$ = 76$\pm$10 (stat) $\pm$10 (syst) nb.
As described in Section 5, CDF later found lower limits
 in the $\pi^+\pi^-$ and $K^+K^-$ channels, implying that most of these $J/\psi + \gamma$ events are from the $\chi_{c1}$ and/or the $\chi_{c2}$.
 LHCb have also observed isolated central $\chi_c \rightarrow J/\psi + \gamma$ states\cite{lhcb}, concluding that most are indeed not $\chi_{c0}$.

\section{$X = \gamma\gamma$}

Exclusive $\gamma\gamma$ production was first proposed in 2001\cite{cdfloi} and was calculated by V.A.Khoze \emph{et al.} \cite{KMRgg}.
It is remarkable in being the only inelastic $pp$ reaction with no produced hadrons or leptons, only two high-$p_T$ $\gamma$-rays.

Exclusive $\gamma\gamma$ production in proton-proton
collisions is dominantly a QCD process based on $g + g \rightarrow \gamma+\gamma$
through quark loops (mainly $u\bar{u}$ and $c \bar{c}$) with another ``screening" gluon exchanged to cancel the color and
allow the protons to emerge intact, see Fig.3. The physics is very similar to that of exclusive Higgs boson production:
$p + p \rightarrow p + H + p$, with gluon fusion through a top-loop. Unlike other exclusive QCD production processes such as
$\chi_c, \chi_b$ and di-jets, these two ($\gamma\gamma$ and $H$) have no produced hadrons, and are therefore very clean.
Measuring exclusive $\gamma\gamma$ production at the LHC will further constrain the uncertainties on the
exclusive Higgs production cross section, of interest for a future PPS/AFP project\cite{royonafppps} with proton detection at $z \sim \pm$420 m.
Independent of that, it probes interesting aspects of QCD and diffraction, as the cross section depends on the unintegrated gluon
distribution in the proton, $G(x_1,x_2,Q^2)$, on Sudakov suppression of additional gluon radiation, and on the rapidity-gap survival
probability $\hat{S}^2$. It provides a bridge between high-$Q^2$ and low-$Q^2$ diffraction.
For exclusive production
of two photons, each with transverse energy~\cite{kine} $E_T^\gamma>5$~GeV and pseudorapidity $|\eta^\gamma|<1.0$, the only predicted cross
section at $\sqrt{s}$ = 1960 GeV~\cite{KMRgg} is $40^{\times \sim 2}_{\div \sim 2}$~fb.
The \emph{fully exclusive} reaction is $ p + p \rightarrow p + \gamma\gamma + p$, requiring detection of both outgoing protons, which 
was not possible in CDF as there were not Roman pots on both sides..
However the fraction of events with an undetected proton dissociation was estimated to be $<$1\%, since rapidity gaps were 
required out to $|\eta_{max}|$ = 7.4

\begin{figure}[!ht]
  \begin{center}
 \includegraphics[width=0.50\textwidth]{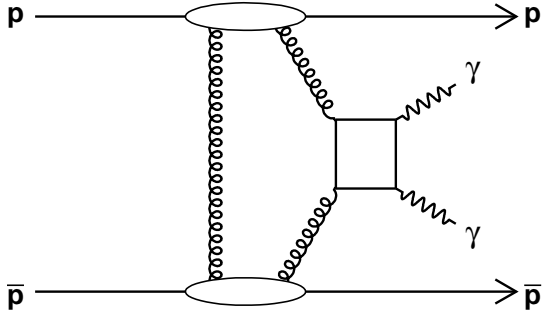}
    \caption{The dominant diagram for central exclusive $\gamma\gamma$ production 
    in $p\bar{p}$ collisions. The primary process is $gg\rightarrow\gamma\gamma$ through quark loops, with a
    screening gluon to cancel the exchanged color.}
  \end{center}
  \label{boxgg}
\end{figure}

Together with the 16 two-track $e^+e^-$ events (Section 2) there were three events that had no tracks\cite{cdfgg1}, and were consistent with 
exclusive two-photon production. For this 
CDF required $|\eta_{\gamma}| < 1.0$ to be sure of excellent tracking efficiency. However possible backgrounds from $\pi^0\pi^0$ or
$\eta^0\eta^0$ were uncertain and \emph{observation} of the process could not be claimed. 
A 95\% C.L. upper limit of $\sigma (|\eta(\gamma)| < 1.0, E_T(\gamma) > 5 $ GeV) of 410 fb at $\sqrt{s}$ = 1.96 TeV was set, 
using $L$ = 532 pb$^{-1}$ of no-pile-up
data.

\begin{figure}[htbp]
  \begin{center}
  \begin{tabular}{cc}
 \includegraphics[width=0.35\textwidth]{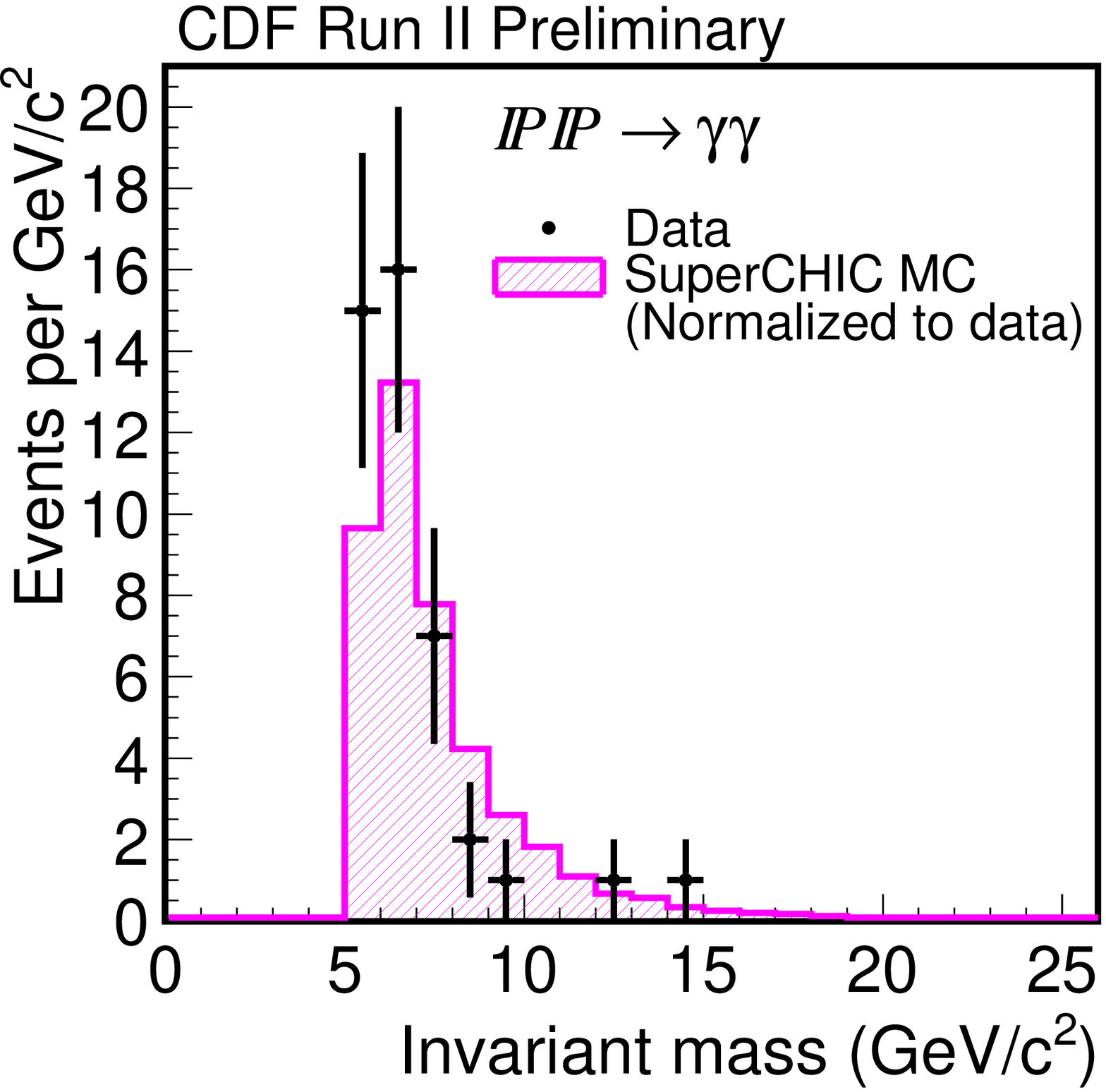}
   &
    \includegraphics[width=0.35\textwidth]{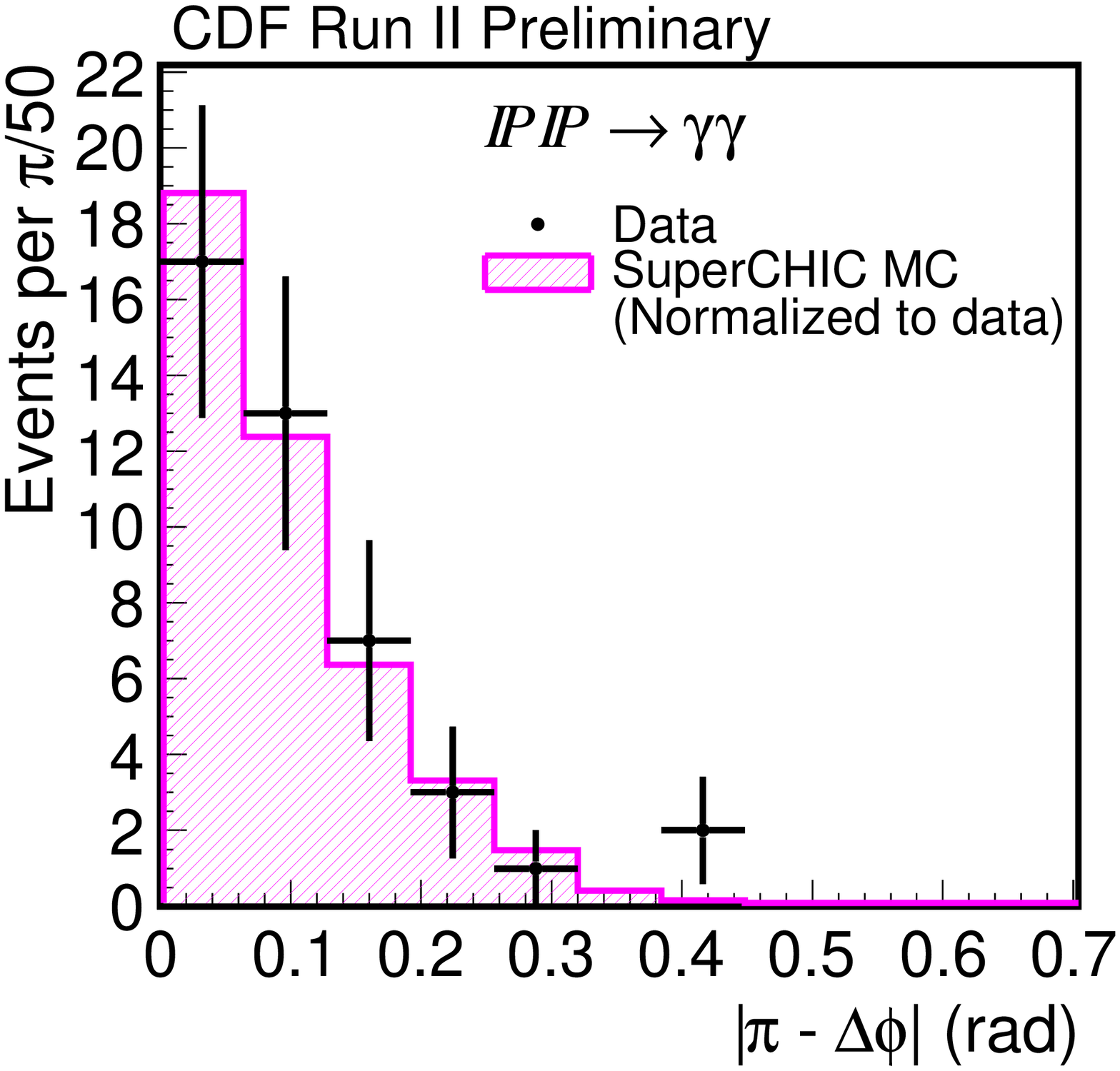}
   \\
   \includegraphics[width=0.35\textwidth]{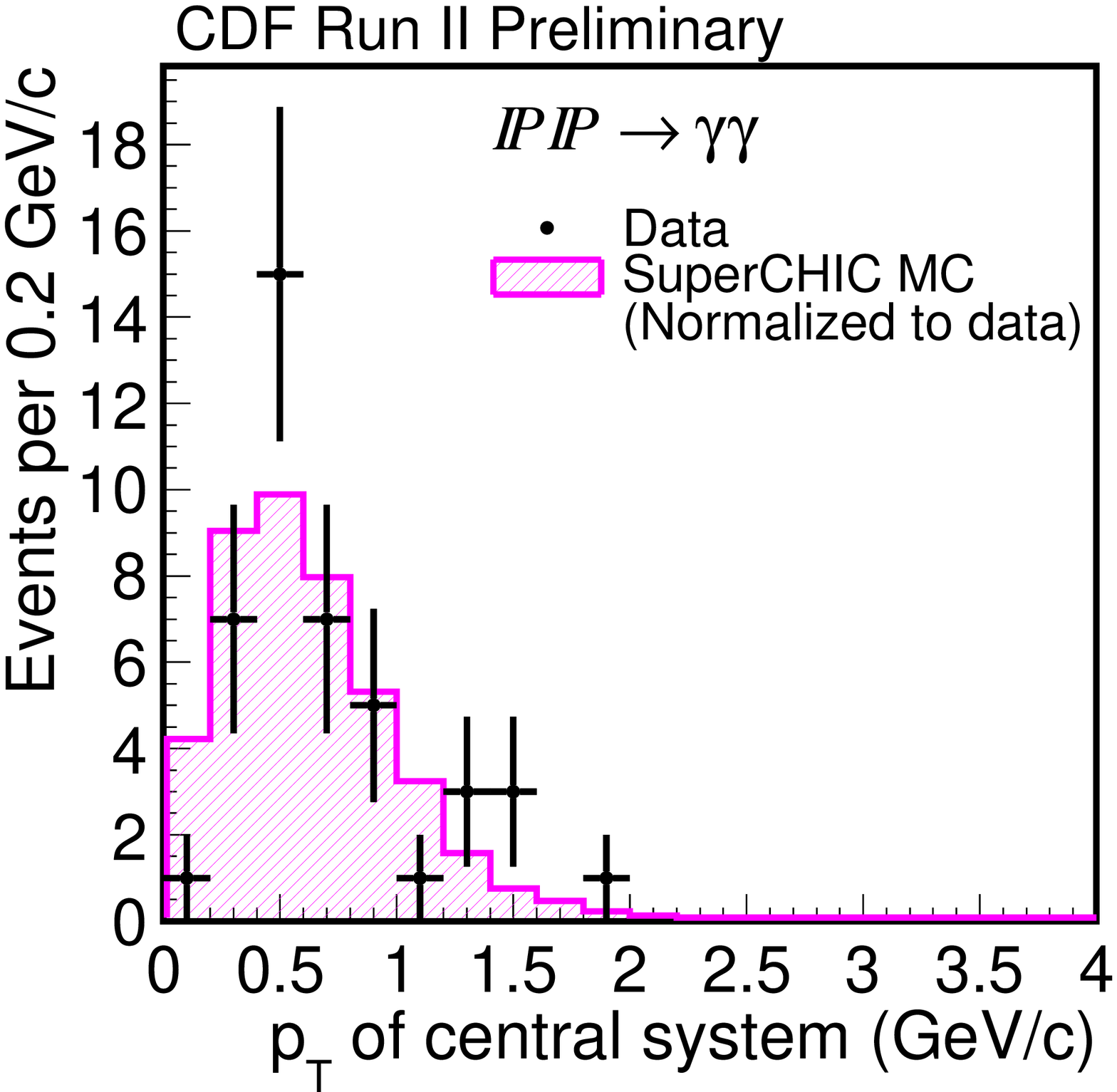}
    &
   \includegraphics[width=0.35\textwidth]{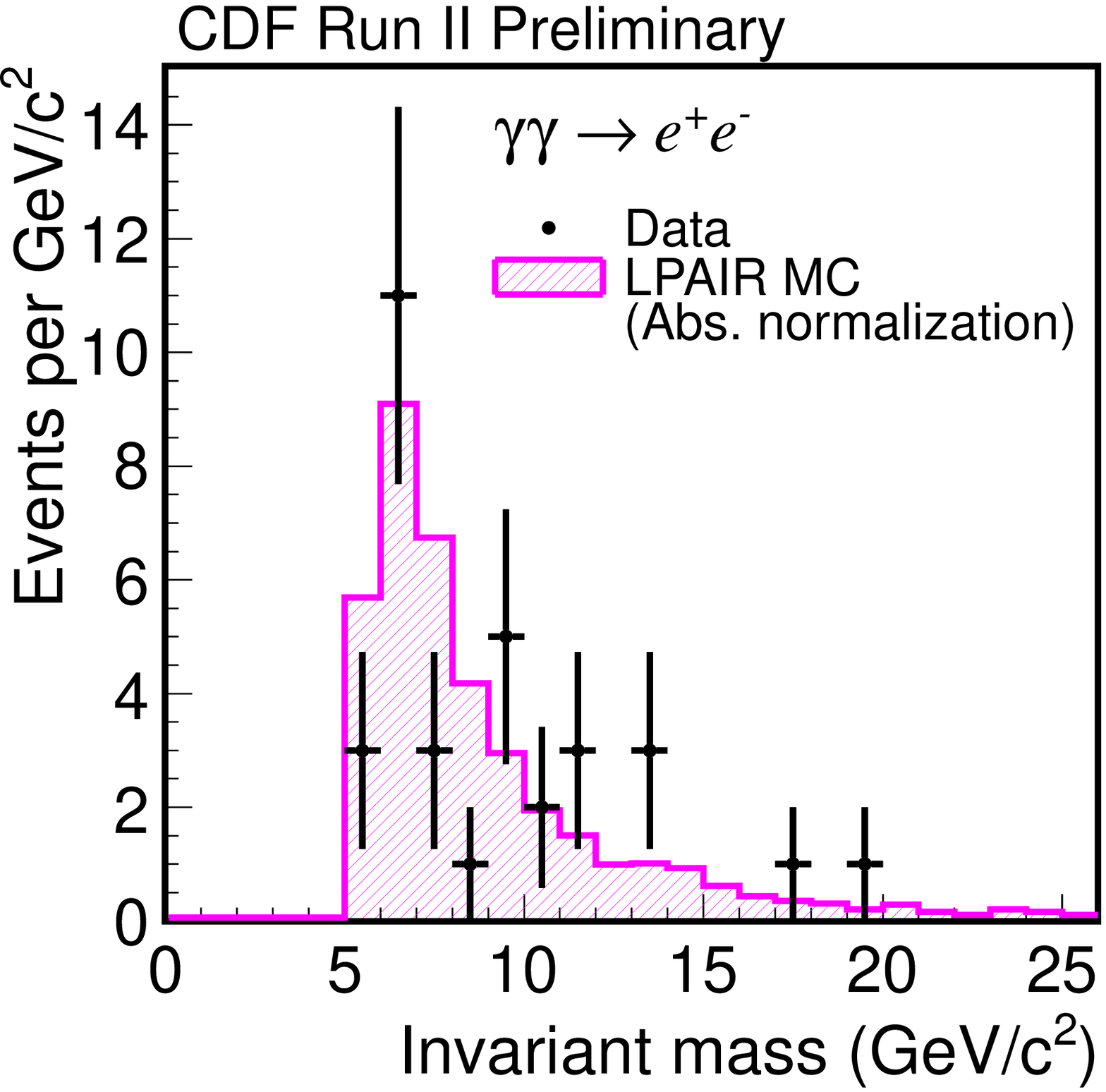}
    \\
 \end{tabular}
  \caption{Two-photon candidates: (a) Invariant mass distribution (b) $|\pi-\Delta\phi|$
distribution (c) $p_T$ distribution of the central system. The invariant mass distribution of $e^+e^-$ is shown in (d).
All error bars are statistical. The MC predictions are normalized to the data for
$\gamma\gamma$, and to the QED calculation~\cite{lpair} for $e^+e^-$.
\label{cdfgg} }
  \end{center}
\end{figure}

CDF was then able to lower the trigger threshold on the
EM clusters from 4 GeV to 2 GeV (2.5 GeV off-line), since this was far above the noise, and as the BSC-1 veto was effective the rate was still acceptable.
This enabled the first observation \cite{cdfgg} of the process, with 43 events having the expected characteristics of two photons 
with $E_T >$ 2.5 GeV.  The full CDF detector, with BSC-3 coverage to $|\eta|$ = 7.4  was required to be empty.
 With the beam rapidity $y$ = 7.64 there is little room for undetected diffractive dissociation. Fig.\ref{cdfgg} shows some kinematic distributions compared with the shapes of the \textsc{superchic} Monte Carlo
predictions\cite{superchic}.

The background from exclusive $\pi^0\pi^0$ production was consistent with zero, as predicted by the Durham Group\cite{durhampiopio}. The cross section, 
$\sigma (|\eta(\gamma)| < 1.0, E_T(\gamma) > 2.5 $ GeV) is about 2.5 pb, consistent with the Durham prediction. The ratio ($\gamma\gamma / e^+e^-$) was, in numbers of candidates,
43/34 with $E_T >$ 2.5 GeV and 3/16 with $E_T >$ 5 GeV, the QED process ($\gamma\gamma \rightarrow \ell^+\ell^-$) having a harder spectrum.

This first (and so far, only) observation of $p + \gamma\gamma + p(\bar{p})$ events \cite{cdfgg} constrains the theory of perturbative exclusive processes.
   From these constraints
we now expect $\sigma(p + H(125) + p) \sim 2^{\times 2}_{\div 2}$ fb at the LHC. This is small, but similar to $\sigma(H) \rightarrow Z Z^*
\rightarrow l_1^+l_1^- l^+_2 l^-_2 \sim$ 1 fb at the LHC (8 TeV) which has been observed. In both cases the continuum background is comparable to the signal, given the
good mass resolution from the lepton or proton measurements. A search for $X = \gamma\gamma$ was made by CMS\cite{cmsgg} based on only 36 pb$^{-1}$, 
and as no candidates were found an upper limit was given compatible with the Durham predictions \cite{KMRgg}.
Unfortunately there has been no LHC running with low enough pile-up in ATLAS and CMS to measure this reaction. (It should be possible, 
in about one week of low pile-up ($\bar{\mu} \sim 1$) running, to make a 10\% meaurement.)

\section{$X = \pi^+\pi^-$}

 Unique to hadron-hadron collisions 
are \pom + \pom $\;$ interactions producing central hadrons between two large rapidity gaps, $\Delta y \gtrsim 4$. CDF has studied exclusive hadron pair production, either direct or from the dacay
of single meson resonances with the allowed quantum
numbers, such as $f_0(600)/\sigma, f_0(980), f_2(1270)$, etc., already seen at lower energies \cite{afsdpe}, an
the $\chi_c$ \cite{cdfmumu} and $\chi_b$ states.
\dpe reactions are expected to be ideal for producing glueballs as the pomeron is glue-dominated.
Of course \emph{any} hadron states can be produced in pairs, including glueballs or states with exotic (non-$q\bar{q}$) quantum numbers.
The Durham group \cite{durhampiopio} has predicted that pairs of isoscalars, such as $\eta^0\eta^0$ should be favored (at high $M(X)$) over
pairs of isovectors, such as $\pi^0\pi^0$, with $\eta' \eta'$ being even more favored as it is thought to have a high gluon content.

Double pomeron exchange, \pom + \pom   or \dpe , was studied at the SPS \cite{kirk}, the Tevatron (fixed target)\cite{gutierrez} and 
ISR \cite{albrowisr, makariev} ($pp$ with $\sqrt{s}$ = 23 - 63
GeV and $\alpha\alpha$ with $\sqrt{s}$ = 126 GeV), but without such large rapidity gaps, allowing some non-\pom   background. At the ISR 
the masses of the central state,
$M(X)$, were limited to about 3 GeV/c$^2$, a good region for meson spectroscopy studies. The quantum numbers of the central state
$X$ are restricted to be $I^G \: J^{PC} = 0^+ \: \mathrm{(even)}^{++}$, and glueball formation is favored as they are isoscalars and the
pomerons are ``glue-rich". Quoting from the PDG \cite{pdg} ``The scalar (isoscalar) mesons are especially important to understand because they
have the same quantum numbers as the vacuum. Therefore they can condense into the vacuum and break a symmetry such as a global chiral
$U(N_f) \times U(N_f)$. The details of how this symmetry breaking is implemented in Nature is one of the most profound problem in
particle physics". So even without discussing glueballs, \dpe  is extremely interesting (see Ref. \citen{acf} for a review).

I now present preliminary results on $p + \bar{p} \rightarrow p(*) + \pi^+\pi^- + \bar{p}(*)$ in CDF. It was possible to trigger on two central
($|\eta| < 1.3$) calorimeter towers with a threshold as low as $E_T$ = 0.5 GeV by vetoing on signals in the BSC counters and the forward plug
calorimeter. There was very little pile-up, easily removed by requiring exactly two tracks (and their associated calorimeter signals)
and no other activity in $|\eta| < 5.9$. This selected events with rapidity gaps of $\Delta \eta > 4.6$ on each side of the $\pi^+\pi^-$ pair. CDF recorded
90 (22) million events at $\sqrt{s}$ = 1960 (900) GeV, the lower energy data in a special 38-hour run they proposed for this
purpose. 

\begin{figure}[!ht]
  \begin{center}
  \includegraphics[width=0.90\textwidth]{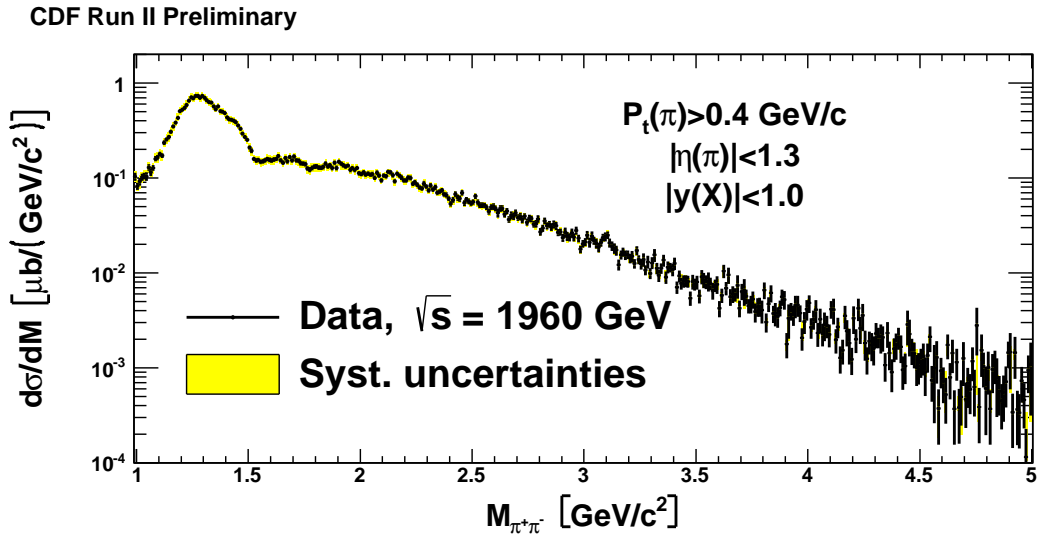}
    \caption{The cross section for exclusive $\pi^+\pi^-$ vs $M(\pi\pi)$ for all $p_T(\pi\pi)$, assuming the hadrons to be pions,
    at $\sqrt{s}$ = 1960 GeV.}
  \end{center}
  \label{masspipi}
\end{figure}

 After all exclusivity and quality cuts CDF had 350,223 (9,349) $h^+h^-$ events. A study of
the time-of-flight shows that $\gtrsim$ 90\% of the events were $\pi^+\pi^-$. Even though the collision time was not known to $\lesssim$ 1 ns, the $h^+$ and
h$^-$ have different momenta and different path lengths, and identification wss possible for $\sim$70\% of track pairs. The acceptance was a function of both
$p_T(\pi\pi)$ and $M(\pi\pi)$ and was calculated for $|y(\pi\pi)| < 1.0$, assuming isotropic decay of $``X" \rightarrow \pi\pi$. As the acceptance was zero
for low $p_T(\pi\pi)$ below $M(\pi\pi)$ = 1000 MeV/c$^2$, CDF presented the cross section, integrated over all $p_T(\pi\pi)$, for higher masses. Fig.
\ref{masspipi}
shows the cross section (assuming pion masses) up to 5 GeV/c$^2$ at $\sqrt{s}$ = 1960 GeV. Features are a large peak, probably both
$f_2(1270)$ and $f_0(1370)$, a break at about 1550 MeV/c$^2$ followed by a smooth, almost-exponential fall off. 
The cross section from threshold, $M(X) = 2m(\pi)$, was presented for $p_T(\pi\pi) > 1$ GeV/c where there is acceptance.
The data show an $f_0(980)$ signal and a cusp (a rapid decrease) at the adjacent $K \bar{K}$
threshold.. A small peak at 3100 MeV/c$^2$ is
consistent with photoproduced $J/\psi$ decaying to $e^+e^-$ (events with muons were excluded). 
The data at $\sqrt{s}$ = 900 GeV look similar, but with
much lower statistics; however the ratio $\sigma(900)/\sigma(1960)$ has some structures \cite{cdfpipi}.
One difference is that the rapidity gaps
extend to $|\eta| = 5.9$ at both energies, while the beam (true) rapidity is $y_{beam} = \mathrm{ln} (\sqrt{s}/m_p)$ = 6.87 and 7.64, so that higher
diffractive masses were included at $\sqrt{s} = $1960 GeV.

In its previous observation of exclusive $\chi_{c0}(3415) \rightarrow J/\psi + \gamma$ \cite{cdfmumu} CDF could not 
resolve the three $\chi_c$ states, and gave
a cross section \emph{assuming} only $\chi_{c0}$.  
CDF then looked for $\chi_{c0} \rightarrow \pi^+\pi^- (K^+K^-)$ which have higher $\chi_{c0}$ branching
fractions, and the mass resolution easily resolves the three states. Fitting the spectrum between 2.5 GeV/c$^2$ and 4.5 GeV/c$^2$ with an
empirical function excluding the $\chi_{c0}$ region, CDF placed upper limits on  its exclusive production: $d\sigma/dy(\chi_{c0})|_{y=0} \lesssim$ 
25 nb (at 90\% C.L.). This is only compatible with the previous observation if $<$ 25\% of the $J/\psi + \gamma$ events were from $\chi_{c0}$. Note
that the branching fractions of $\chi_{c1} (\chi_{c2}) \rightarrow J/\psi+\gamma$ are 30$\times$ (17.4$\times$) higher, and they may still be suppressed in
production, as expected \cite{kmr3}.

CDF has more data available for other channels, and $X = K^+K^-, p \bar{p}, \pi^0\pi^0, \eta^0\eta^0, \eta^0\eta'$, are being studied, 
as well as $\eta' \eta'$ production with four
photon showers and 0, 1, or 2 pairs of charged pions. The Durham Group\cite{durhampiopio} predicts a hierarchy, 
with $\sigma(\eta(')\eta(')) > \sigma(\pi^0\pi^0)$ at large $M(X)$.
This is due to the $J_z = 0$ selection rule, which states that the CEP of any pair of flavor non-singlet mesons vanishes when the fusing gluons are in
the $J_z = 0$ state. For flavor singlet mesons such as $\eta(') \eta(')$ there are additional ladder diagrams for which the $J_z = 0$ amplitudes
do not vanish. Possible gluon components to these mesons can also enhance the cross section.
 This would be very interesting; there is data but the analysis is still ongoing.  

\begin{table}
\begin{center}
\begin{tabular}{|c|c|c|c|c|c|c|c|}
\hline
\hline
 Reaction  & Ref.  & $L_{eff}$ & \# ev. & $|\eta_{max}|$ & $E_T(min) $ & $M(X)$    & $\sigma$ \\
    ($X$) &       & pb$^{-1}$ &        &                & GeV         &  GeV/c$^2$ &  \\
\hline
  $e^+e^-$ & \cite{cdfee} & 45.8      & 16     & 2.0 & 5   &  & (1.6$\pm$0.6) pb \\
$\ell^+\ell^-$           & \cite{cdfz}      & $~\sim$435          &    8    &  4      &    & 40-80 & (0.24$\pm$0.13) pb \\
   $e^+e^-$    &  \cite{cdfgg}     & 34   &  75.5          &   1.0         &   2.5            & 5 - 20    &  (2.88$\pm$0.85) pb            \\
    $Z$       & \cite{cdfz}      &    $\sim$435       &   0     &    4    &     & 82-98 & $<0.25$ pb (95\%)$^1$\\
 $\mu^+\mu^-$   & \cite{cdfmumu}      &   139$\pm$8        & 77       &   0.6     &  1.4   & 3.0 - 4.0 & (2.7$\pm$0.5) pb\\
  $J/\psi$       & \cite{cdfmumu}     &   139$\pm$8   &   286    &  0.6          & 1.4  &     & (3.92$\pm$0.58) nb$^1$\\
$\psi(2S)$       & \cite{cdfmumu}    & 139$\pm$8      & 39       & 0.6           & 1.4  &     & (0.53$\pm$0.14) nb $^1$\\
$\chi_c$         & \cite{cdfmumu}     & 139$\pm$8     & 65       & 0.6           & 1.4  &     & (76$\pm$14)   nb    $^2$ \\
$\chi_{c0} (\pi\pi)$ & \cite{cdfpipi} & 1.16$\pm$0.12 & $< 49.3$ & 1.0           & 0.4  &     & $<$ 37.6  nb (90\%)\\ 
$\chi_{c0}(KK) $ & \cite{cdfpipi}     & 1.16$\pm$0.12 & $< 34.6 $& 1.0           & 0.4  &     & $<$ 24.8  nb (90\%)\\
$\gamma\gamma$   &  \cite{cdfgg1}    &   45.8         & 3        & 1.0           & 5    &     &  $<$ 410 fb (95\%) \\
$\gamma\gamma$   & \cite{cdfgg}     &   75.5         & 43       & 1.0           & 2.5  &     &  2.48$\pm$0.65 pb    \\
   $JJ$          &  \cite{goulianos} &  310           &    -     &   2.5         &  25  &     &  4.84$^{+4.2}_{-3.4}$ pb$^3$        \\
\hline
 \end{tabular}
 \\[4.0mm]
 Table 1 : Summary of cross sections and limits from CDF CEP. 
 Statistical and systematic uncertainties have been 
 combined in quadrature. Note 1: $d\sigma/dy$ at $y = 0$. 
 Note 2: Assuming all $\chi_{c0}$. Note 3: See Ref.\citen{goulianos} for other values.
 \end{center}
  \label{sigmas}
\end{table}

\section{Other CDF \dpe results}

For completeness I mention the observation of exclusive di-jets reported in this issue by Goulianos \cite{goulianos}, with one value among several being 
included in the summary Table 1, and also a measurement of \emph{inclusive} \dpe \cite{incldpe}. This used a triplet of Roman pots to measure diffractively
scattered antiprotons, selecting inelastic events with no pile-up, and finding
the fraction of events that had a rapidity gap on the opposite (proton) side.
The fractional momentum loss $\xi_p$ of the unobserved proton was calculated from the relation:

\[ \xi^X_p = \frac{1}{\sqrt{s}} \sum_{i=1}^n E^i_T e^{\eta^i}, \]

where $E_T^i$ and $\eta^i$ are the transverse energy and pseudorapidity of a particle, and the sum is carried out over all particles except the proton. 
For events with an antiproton with $0.035 < \xi_{\bar{p}} < 0.095$ the fraction that have a proton with $\xi_p < 0.02$ is 0.194$\pm$0.012.  

\section{Summary}

While the physics program at the Tevatron focused on the highest $Q^2$ regime (top-quark, jets, supersymmetry and 
Higgs searches, etc.) diffractive and central exclusive physics was a significant part of the QCD studies. 
With special forward detectors and triggers CDF could take advantage of occasional low pile-up running to observe the
several new and simple reactions as described above, and summarized in Table I. Especially for \pom + \pom $\rightarrow$ hadrons with $M(X) \lesssim$ 5 GeV/c$^2$, the Tevatron is
in principle as good as the LHC, and better than lower $\sqrt{s}$ machines, so it is a pity that it was closed without more time being dedicated to this physics.

\section*{Acknowledgments}

I would like to thank all members of the CDF Collaboration, especially my young colleagues who worked hard to extract the physics from the data:
 L.Zhang, A.Hamilton, E.Nurse, E.Br\"{u}cken, M.Zurek, A.Swiech, D.Lontkowski, and I.Makarenko.


\begin{thebibliography}{0}
\bibitem{kine} A cylindrical coordinate system is used with the $z$-axis along the 
proton beam direction; $\theta$ is the polar angle and $\phi$ is the azimuthal angle.
pseudorapidity is defined as $\eta= -\ln \tan(\theta/2)$,
transverse momentum as 
$p_T =|p| \sin\theta$, and transverse energy as $E_T = E \sin\theta$.
\end{thebibliography}
\end{document}